\documentclass[prl,reprint,superscriptaddress]{revtex4-1}

\usepackage{amsmath,amssymb}
\usepackage{verbatim}
\usepackage{graphicx}
\usepackage{hyperref}
\usepackage{color}
\usepackage{mathrsfs}

\newcommand {\cL}{{\cal L}}

\newcommand {\cN}{{\cal N}}

\def\a{\alpha}
\def\b{\beta}
\def\c{\chi}
\def\d{\delta}
\def\e{\epsilon}

\def\l{\lambda}
\def\m{\mu}
\def\n{\nu}
\def\o{\omega}

\def\r{\rho}
\def\s{\sigma}
\def\t{\tau}

\newcommand{\ve}{\varepsilon}

\newcommand{\pa}{\partial}

\newcommand{\vf}{\varphi}

\newcommand{\be}{\begin{equation}}
	\newcommand{\ee}{\end{equation}}
\newcommand{\bea}{\begin{eqnarray}}
	\newcommand{\eea}{\end{eqnarray}}
\newcommand{\non}{\nonumber}
\newcommand{\ba}{\begin{array}}
	\newcommand{\ea}{\end{array}}

\def\double #1{#1{\hbox{\kern-2pt $#1$}}}

\newcommand{\bsubeq}{\begin{subequations}}
	\newcommand{\esubeq}{\end{subequations}}

\newcommand{\rd}{\mathrm d}
\newcommand{\eps}{\varepsilon}

\newcommand{\bbD}{\mathbb D}

\def\ft#1#2{{\textstyle{\frac{\scriptstyle #1}{\scriptstyle #2} } }}

\begin{document}

\title{Gauss-Bonnet supergravity in six dimensions}

\author{Joseph Novak}
\email{joseph.novak@aei.mpg.de}
\affiliation{Max-Planck-Insitut f\"ur Gravitationsphysik (Albert-Einstein-Institut)
Am M\"uhlenberg 1, DE-14476 Potsdam, Germany}

\author{Mehmet Ozkan}
\email{ozkanmehm@itu.edu.tr}
\affiliation{Department of Physics,
Istanbul Technical University,
Maslak 34469 Istanbul,
Turkey}

\author{Yi Pang}
\email{Yi.Pang@aei.mpg.de}
\affiliation{Max-Planck-Insitut f\"ur Gravitationsphysik (Albert-Einstein-Institut)
Am M\"uhlenberg 1, DE-14476 Potsdam, Germany}

\author{Gabriele Tartaglino-Mazzucchelli}
\email{gabriele.tartaglino-mazzucchelli@kuleuven.be}
\affiliation{Instituut voor Theoretische Fysica, KU Leuven,
Celestijnenlaan 200D, B-3001 Leuven, Belgium}

\date{\today}
	
%	\preprint{}

\begin{abstract}
The supersymmetrization of curvature squared terms is important in the study of the low-energy limit of compactified 
superstrings where a distinguished role is played by the Gauss-Bonnet combination, which is ghost-free.
In this letter, we construct its off-shell ${\cal N} = (1, 0)$ supersymmetrization in six dimensions for the first time.
By studying this invariant 
together with the supersymmetric Einstein-Hilbert term we confirm and extend known results of the $\alpha'$-corrected string 
theory compactified to six dimensions. Finally, we analyze the spectrum about the ${\rm AdS}_3\times{\rm S}^3$ solution.

\end{abstract}
	
%	\pacs{??? ... ???}

\maketitle
\allowdisplaybreaks

\textit{Introduction.}---Six-dimensional (6D) supergravities 
are of considerable interest for numerous 
reasons that are rooted 
in their connection to superstring theory. This connection often guarantees improved quantum behavior for such theories. 
For instance, 6D anomaly-free matter-coupled models are known to arise from T$^4$ or K3 compactification of heterotic or 
type II string theories \cite{Green:1984bx,Walton:1987bu}, see also \cite{Nishino:1986dc-1997ff}. 
Another noteworthy model, 
the Salam-Sezgin model \cite{Salam:1984cj}, which has been found to have an M/string theory origin 
\cite{Cvetic:2003xr}, admits a unique supersymmetric ${\rm M}_4 \times {\rm S}^2$ 
vacuum \cite{Gibbons:2003di} that may have 
interesting phenomenological applications 
\cite{Maeda:1984gq-1985es,Gibbons:1986xp,Halliwell:1986bs,Aghababaie:2002be}.

Supergravity models in 6D also play an important role in the AdS$_3/$CFT$_2$ correspondence and BTZ black hole 
microstate counting since the ungauged theory 
accommodates a supersymmetric ${\rm AdS}_3 \times {\rm S}^3$ solution \cite{deBoer:1998kjm, David:2002wn}. 
However, precision tests of holography often require knowledge of supersymmetric higher order curvature invariants, 
motivating the need for their construction.

Higher order curvature terms are also important in string theory where the corrections take the form of an 
infinite series constrained by the on-shell supersymmetry order by order in the string tension $\alpha'$. 
Upon reduction to six dimensions, where an 
off-shell formulation of supergravity is available, exact higher curvature invariants can be systematically constructed. 
The leading corrections 
come from curvature squared terms in which the Gauss-Bonnet (GB) combination 
$ R^{\mu\nu\rho\s} R_{\mu\nu\rho\s} - 4 R^{\mu\nu} R_{\mu\nu} + R^2=6R_{[\mu\nu}{}^{\mu\nu} R_{\rho\s]}{}^{\rho\s}$
 is singled out as it is ghost free, and its equations of motion are second order in 
derivatives \cite{Zwiebach:1985uq,Deser:1986xr}. These features facilitate the study
of exact solutions and significantly 
simplifies the computation of physical quantities.

The purely gravitational higher curvature terms are related by supersymmetry to 
contributions depending on $p$-forms. These terms, that have not yet been systematically analyzed in the literature, 
play an important role in understanding the moduli in compactified string theory and the low-energy description of 
string dualities; see, e.\,g., \cite{Antoniadis:1997eg,Antoniadis:2003sw,Liu:2013dna}.

The construction of the GB supergravity invariant can be achieved using off-shell techniques for
${\rm D}\leq 6$. Their construction in 4D and 5D were presented in \cite{N=1-GB,Butter:2013lta}
and \cite{OP131,OP132, Butter:2014xxa}, respectively, while for the 6D
case only partial results were obtained 30 years ago \cite{BSS1,BSS2,Nishino:1986da,BR}.
In this work we complete the construction of the 6D off-shell $\cN=(1,0)$
GB term utilizing the techniques of \cite{BSVanP,BKNT16, BNT-M17}.

A major advantage of an off-shell formulation is that the supersymmetry transformations 
do not receive higher-order corrections.
It also allows one to easily combine separate Lagrangians. 
In fact, by adding the off-shell supersymmetric Einstein-Hilbert term of \cite{BSVanP} to our GB invariant we obtain an off--shell 
completion of the 6D Einstein-Gauss-Bonnet supergravity action originally derived in \cite{Liu:2013dna} by using the 
heterotic/IIA duality reduced to 6D $\cN=(1,1)$, see also \cite{Romans:1985tw}, 
and by truncating to the NSNS $\cN=(1,0)$ sector.
We conclude by analyzing the $\alpha'$-corrected spectrum of fluctuations
around the supersymmetric ${\rm AdS}_3\times{\rm S}^3$ solution.

\textit{New supergravity invariants.}---To begin with, 
we briefly summarize the field content of
the standard Weyl multiplet of $\cN = (1,0)$ conformal
supergravity in six dimensions \cite{BSVanP}.
This consists of $40+40$ bosonic and
fermionic off-shell degrees of freedom.
In the following we use the conventions of
\cite{Coomans:2011ih,Bergshoeff:2012ax}
with the exception of a sign difference in the parity transformation.
We denote spacetime indices
by $\mu,\nu,\ldots$, tangent space indices by $a,b,\ldots$ and
$\mathrm{SU}(2)$ indices by $i,j,\ldots$
Among the bosonic fields
are the sechsbein $e_\mu{\!}^a$, $\mathrm{SU}(2)$ gauge fields
$V_\mu{}_i{}^j$ and a gauge field $b_\mu$ associated with
dilatations. Furthermore, there are two composite bosonic gauge
fields, namely the spin connection ${\omega}_\mu{\!}^{ab}$ and the gauge
field ${f}_\mu{\!}^a$ associated with conformal boosts.  In
addition to the gauge connections, the bosonic content of the standard Weyl multiplet comprises of
an anti-self-dual tensor
$T_{abc}^-$
and a real scalar $D$
as covariant
matter fields.
The fermions consist of the gravitini
$\psi_{\mu i}$ that are the gauge fields of $Q$-supersymmetry, a composite
gauge field ${\phi}_{\mu}{\!}^i$ associated with $S$-supersymmetry,
and a chiral fermion $\c^i$.
In what follows, we will suppress the fermionic terms and
we will refrain from discussing the superconformal transformations
of the various fields in any detail, which are
summarized in \cite{Coomans:2011ih}.
	
To describe curvature squared terms,
we will make use of a variant 40+40 multiplet of conformal supergravity,
known as the dilaton-Weyl
multiplet \cite{BSVanP}.
This	is obtained by coupling the standard Weyl multiplet to a tensor multiplet.
The independent connections remain the same as in the standard Weyl multiplet
but the covariant matter fields
are exchanged with
those of the tensor multiplet, which include a scalar field $\s$,
a gauge two-form $B_{\mu\nu}$
and a spinor $\psi^i$
among its component fields. Upon doing so the bosonic covariant fields of the
standard Weyl multiplet are then expressed as\bsubeq
\label{dil-Weyl}
\bea T_{abc}^- &=& \ft{1}{2 \s} H_{abc}^- , \\
D &=& \ft{15}{4 \s} \big( {\mathscr{D}}^a {\mathscr{D}}_a \s + \ft{1}{5} R \s + \ft{1}{3} T^-_{abc} H^{abc} \big)
+ {\rm f.t.},
~~~~~~
\eea\esubeq
where  ``${\rm f.t.}$'' stands for neglected fermionic terms and
$H_{abc}^-$ denotes the anti-self-dual part of the three-form field strength
$H_{abc} = 3 e_a{}^\mu e_b{}^\nu e_c{}^\rho \partial_{[\mu} B_{\nu\rho]}$.
Here we have made use of the covariant derivative
\be \label{hatDder}
{\mathscr{D}}_a =e_a{}^\mu\big( \partial_\mu
- \ft12 \omega_\mu{}^{bc} M_{bc}
- b_\mu \bbD
- V_\mu{}_i{}^{j} U_{j}{}^i
\big),
\ee
where $M_{ab}$, $\bbD$, and $U_i{}^j$ are the Lorentz, dilatation, and SU(2) generators, respectively.
The Lorentz curvature  is given by
$R_{\mu\nu}{}^{cd}=R_{\mu\nu}{}^{cd}({\o}):=2\pa_{[\mu}{\o}_{\nu]}{}^{cd}+2{\o}_{[\mu}{}^{ce}{\o}_{\nu]}{}_e{}^d$,
where ${\o}_{\mu}{}^{cd}=\omega(e)_{\mu}{}^{cd}+2e_\mu{}^{[c}b^{d]}$
is the Lorentz connection and  $\omega(e)_{\mu}{}^{cd}$  is the usual torsion-free spin connection.
The dilatation connection $b_\mu$ is pure gauge \cite{BSVanP}, and 
we will always consider it
set to zero.
Besides the Riemann tensor $R_{ab}{}^{cd}:=R_{ab}{}^{cd}(\o)$,
we also use the tensors
$R_{a}{}^{b}=R_{ac}{}^{bc}$ and $R=R_{ab}{}^{ab}$,
which coincide with the
Ricci and scalar curvature tensors, respectively.
Note that, by using the mapping \eqref{dil-Weyl}, every invariant involving a coupling to the
standard Weyl multiplet can be directly converted to one in terms of the dilaton-Weyl multiplet.

In three, four, and five dimensions with eight supercharges, a special role is
played by the linear multiplet action principle coupled to supergravity \cite{BF4,BF5,BF3}.
It schematically represents an action based on the product of a linear and
a vector multiplet and describes supersymmetric extensions of $B_{d-2}\wedge F_2$ invariants
in $d$-dimensions with $F_2$ a closed two-form and $B_{d-2}$ an unconstrained $(d-2)$-form.
Although this action principle also exists for $(1,0)$ superconformal symmetry in six dimensions
and, in fact, was the main building block for superconformal invariants in \cite{BSVanP},
there is also another possibility.
As shown in \cite{BKNT16}, one can generate an action principle
that is schematically the product of the tensor multiplet and a four-form
multiplet, which is the supersymmetric extension of $B_2\wedge H_4$. 
In components the associated Lagrangian takes the form
\be
e^{-1}\cL_{B_2\wedge H_4} =
\ft{1}{4}B_{ab}C^{ab}
- \ft{1}{4}\s C
+ {\rm f.t.}
\label{BH-action}
\ee
Here $C^{ab}=-\frac{1}{4!}\ve^{abcdef}H_{cdef}$ is the Hodge dual of a closed four-form,
$\rd H_4=0$, and $C$ is a real scalar field.
The fields
$C_{a b}=\frac{1}{12} \bar{Q}_{ i}\gamma_{[a}Q_{ j} B_{b]}^{ij}$ and
$C=\frac{1}{12} \bar{Q}_{i}\gamma^{a}  Q_{j} B_a^{ij}$ 
of dimension 4 
are descendant components
of the four-form multiplet \cite{AriasLinchRidgway}
which are defined in terms of a primary field  $B_{a}^{ij} = B_{a}^{(ij)}$ of 
dimension 3 (see also \cite{BKNT16,BNT-M17}),
and $Q^i$ are the $Q$-supercharges.
The density formula \eqref{BH-action},
which extends the one first introduced to describe the rigid supersymmetric Yang-Mills action
\cite{HoweSierraTownsend},
is the building block for constructing the curvature squared invariants in this letter.

Recently, a particular composite four-form multiplet defined solely using the fields of the standard Weyl multiplet
has been constructed in \cite{BKNT16}. It was used to describe an 
$\cN = (1,0)$ conformal supergravity action \cite{BKNT16, BNT-M17}.
The primary dimension 3 field of this multiplet has the form
$B_a^{ij} =\frac{1}{4}T^-_{a b c}\, F^{b c\, i j}+{\rm f.t.}$,
where
$F_{\mu\nu}{}^{kl}:=2\pa_{[\mu}V_{\nu]}{}^{kl}-2V_{[\mu}{}^{p(k} V_{\nu]}{}_p{}^{l)}$
denotes the SU(2) curvature.
The $C$ and $C^{ab}$ components of this composite multiplet
were worked out in \cite{BNT-M17}.
Plugging these results into \eqref{BH-action}
gives the new invariant 
\begin{widetext}
\bea
e^{-1}\cL_{{\rm new}} &=&
\ft{1}{32}
\Big\{\,
\s C_{ab}{}^{cd}  C_{cd}{}^{ab}
- 3\s  F_{ab}{}^{ij} F^{ab}{}_{ij}
+ \ft{4}{15}\s  D^2
- 8\s  T^{- dab}\big(\mathscr{D}_d \mathscr{D}^c T^-_{ab c}
+ \ft12 R_d{}^c T^-_{abc}
\big)
+ 4\s  ( {\mathscr{D}}_c T^{- abc} )  {\mathscr{D}}^d T^-_{abd}
~
\non\\
&&~~~~~
+ 4\s  T^{- abc} T^-_{ab}{}^d T^{- ef}{}_c T^-_{efd}
- \ft{8}{45} H_{abc} T^{- abc} D
+2 H_{abc} C^{ab}{}_{de} T^{- cde}
+ 4  H_{abc} T^-_d{}^{ab} {\mathscr{D}}_e T^{- cde}
\non\\
&&~~~~~
- \ft{4}{3} H_{abc} T^{- dea} T^{- bcf} T^-_{def}
-\ft{1}{4} \eps^{abcdef} B_{ab} \big(
C_{cd}{}^{gh} C_{efgh} - F_{cd}{}^{ij} F_{ef}{}_{ij}
\big)
\Big\}
+{\rm f.t.}
\label{newInvariant}
\eea
\end{widetext}
Here $C_{ab}{}^{cd}$ is the Weyl tensor and
\eqref{dil-Weyl} holds in \eqref{newInvariant}.
The fermionic extension of \eqref{newInvariant} will appear in \cite{BNOPT-M17}.

A supersymmetric extension of the Riemann curvature squared term
was constructed in \cite{BSS1,BSS2,Nishino:1986da,BR}.
This was based on the action for a Yang-Mills multiplet coupled to conformal supergravity \cite{BSVanP}
and exploiting the feature that
in the gauge
\be 
\s = 1, \quad b_\mu = 0,\quad \psi^i=0,
\label{gaugeSigma=1}
\ee
the dilaton-Weyl multiplet can be mapped to
a Yang-Mills vector multiplet taking values in the 6D Lorentz algebra \cite{BSS1}.
It turns out that both the Yang-Mills and the supersymmetric Riemann squared invariants
can be constructed by using the $B_2\wedge H_4$ density formula together with
appropriately chosen composite four-form multiplets.
In superspace these were described in Sec. 6 of \cite{BNT-M17} from which their component actions can be readily 
obtained.
For the purpose of this letter, it is enough to present the bosonic part of the Riemann squared invariants in the gauge
\eqref{gaugeSigma=1}, which takes the form
\bea
&&e^{-1}\cL_{{\rm{Riem}}^2}
=
-4F^{ab}{}_{ij}F_{ab}{}^{ij}
+R^{ab}{}^{cd}(\o_-)R_{ab\,cd}(\o_-)
\non\\
&&~~~~~~
-\ft{1}{4}\eps^{abcdef}B_{ab}\,R{}_{cd}{}^{gh}(\o_-)R_{ef\,gh}(\o_-)
+{\rm f.t.}~~~
\eea
Here $R{}_{ab}{}^{cd}(\o_-)$
is  the torsionful Lorentz curvature defined 
in terms of the modified connection
\bea \label{TFconecOmega}
{\o_\pm}{}_\mu{}^{cd}
:=
{\o}_\mu{}^{cd}
\pm\ft{1}{2}e_\mu{}^aH_a{}^{cd}
\eea
and it is such that
\be
R{}_{ab}{}^{cd}(\o_\pm)
=
R_{ab}{}^{cd}
\pm\mathscr{D}_{[a}H_{b]}{}^{cd}
-\ft{1}{2}H_{e[a}{}^{[c}H_{b]}{}^{d]e}.
\label{TorsionflRiemann}
\ee
As noted in \cite{BSS1},
it is  straightforward to restore a general gauge and undo the condition \eqref{gaugeSigma=1}.

Now that we have described the new curvature squared invariant,  
we are ready to describe an off-shell extension of the 6D Gauss-Bonnet  combination. 
It suffices to take
the following combination of the Riemann squared and the new invariant
\bea
\cL_{{\rm GB}}=
-3\cL_{{\rm Riem}^2}
+128\cL_{{\rm new}}.
\label{GBaction}
\eea
For the applications in this letter we will use the Gauss-Bonnet invariant in the gauge \eqref{gaugeSigma=1}.
Making use of \eqref{dil-Weyl} and \eqref{gaugeSigma=1}
together with \eqref{TorsionflRiemann}, \eqref{GBaction} takes the form
\bea
&&e^{-1}{\cal L}_{\rm GB}= 
6R_{[\m\n}{}^{\m\n} (\o_+) R_{\r\s]}{}^{\r\s}  (\o_+) 
+\ft23 R(\o_+) H^2  \non\\
&&~~ - 4 R^{\m\n}(\o_+) H_{\m\n}^2+ 4 R_{\m\n\r\s}(\o_+) H^{\m\r\a} H^{\n\s}{}_\a    \non\\
&&~~ + \ft19 (H^2)^2 - \ft23 H^4 +\e^{\m\n\r\s\l\t} B_{\m\n} F_{\r\s}{}^{ij} F_{\l\t\,ij} \non\\ 
&&~~ +\ft14\e^{\m\n\r\s\l\t} B_{\m\n} R_{\r\s}{}^{\a}{}_{\b}(\omega_+) R_{\l\t}{}^{\b}{}_{\a}(\omega_+)
+{\rm f.t.},
\label{GBA}
\eea
where
$H^4:=H_{\m\n\s}H_{\r\l}{}^{\s}H^{\m\r\d}H^{\n\l}{}_{\d}$,
$H^2_{\mu\nu}:=H_{\mu}{}^{\rho\s}H_{\nu\rho\s}$
and
$H^2:=H_{\mu\nu\rho}H^{\mu\nu\rho}$. 

It is important to note that the $B$-field dependence of the 
supersymmetric GB invariant cannot be captured solely by a torsionful connection.
This explains the previous unsuccessful
attempts at the supersymmetrization of the 6D GB
action \cite{BSS1,BSS2} where only the first and last two terms in \eqref{GBA} appeared
in the bosonic part of the invariant.
It is worth remarking that  the off-shell GB action
 does not contain higher-order kinetic terms for the 
massless supermultiplet of the two-derivative theory
and the kinetic term for the  $\mathrm{SU}(2)$ gauge fields $V_\m{}^{ij}$ drops out.

\textit{On-Shell Einstein-Gauss-Bonnet Supergravity.}---Now let us study a certain linear combination of
the Einstein-Hilbert (EH) and GB invariants which we refer to as Einstein-Gauss-Bonnet supergravity.
In contrast, to the GB invariant which is based solely on the dilaton-Weyl multiplet, the Einstein-Hilbert
invariant requires a compensating multiplet, which we choose
to be the linear multiplet \cite{BSVanP,Coomans:2011ih}.
It consists of an
SU(2) triplet of scalars $L_{ij}$, a constrained vector field $E_{a}$, and an SU(2) Majorana spinor $\vf^i$. Adopting the
gauge \eqref{gaugeSigma=1}
together with the ${\rm SU}(2)\to{\rm U}(1)$ gauge fixing conditions
\be
L_{ij}=\ft1{\sqrt2}\delta_{ij}L,\quad
V_{\m}^{ij}=V'_{\m}{}^{ij}+\ft12\delta^{ij}V_{\m},
\label{GaugeFixing-SU(2)}
\ee
where $L^2 = L_{ij} L^{ij}$,
the EH action takes the form \cite{Bergshoeff:2012ax}
\bea
&&e^{-1}{\cal L}_{\rm EH}=LR+L^{-1}\partial_{\m}L\partial^{\m}L-\ft1{12}LH_{\m\n\r}H^{\m\n\r}\non\\
&&\qquad\quad+2LV'^{ij}_{\m}V'^{\m}_{ij}-\ft12L^{-1}E^{\m}E_{\m}+\sqrt{2}E^{\mu}V_{\mu}.~~~
\label{eha}
\eea
The off-shell Einstein-Gauss-Bonnet supergravity is defined by the Lagrangian
\bea
2\kappa^2\cL &=& \cL_{\rm EH} + \ft{1}{16} \a^\prime  \cL_{\rm GB}.
\label{GB-0}
\eea
The solution $V_\m{}^{ij} = E_\m = 0$ is consistent with the equations of motion for the  fields $(V_\m{}^{ij}, E_\m)$.
Setting $L = e^{-2\upsilon}$ and pulling out most of the dependence on the three-form $H_3$,  
we obtain the Lagrangian for the on-shell Einstein-Gauss-Bonnet
supergravity
\bea
&&2\kappa^2e^{-1} \cL = e^{-2\upsilon} [R + 4 \partial_\m \upsilon \partial^\m \upsilon  - \ft1{12}H_{\m\n\r} H^{\m\n\r} ] \non\\
&&~ + \ft{1}{16} \a^\prime \Big[6R_{[\m\n}{}^{\m\n} R_{\r\s]}{}^{\r\s}
+ \ft{1}{6} R H^2
- R^{\m\n} H_{\m\n}^2
+ \ft{5}{24} H^4
\non\\
&&\qquad
+\ft12 R_{\m\n\r\s} H^{\m\n\l} H^{\r\s}{}_{\l} 
+ \ft{1}{144} (H^2)^2 -\ft18 (H^2_{\m\n})^2 
\non\\
&&\qquad
+\ft14\e^{\m\n\r\s\l\t} B_{\m\n} R_{\r\s}{}^{\a}{}_{\b}(\omega_+) R_{\l\t}{}^{\b}{}_{\a}(\omega_+)\Big]
+{\rm f.t.}
~~~~
\label{OnShellGB}
\eea
This result has a remarkable feature.
In \cite{Liu:2013dna}, it was conjectured that for the type II string, the $B$-field dependence in
$R^4$ corrections is nearly completely captured
in terms of the torsionful Riemann tensor
(\ref{TorsionflRiemann}) (except for the CP-odd sector). 
The claim was further studied by fixing the one-loop four-derivative
corrections in six dimensions by means of a K3 reduction of type IIA 
and requiring that the dyonic string remains a solution, 
as well as the duality of this model to heterotic strings compactified
on T$^4$.
Here we provided an alternative derivation of the four-derivative corrections by the exact
supersymmetrization of the curvature squared invariants. Our result for the on-shell Einstein-Gauss-Bonnet
supergravity precisely
matches \cite{Liu:2013dna}, thereby providing strong evidence for the conjecture 
put forward there.

\textit{$AdS_3\times S^3$ solution and spectrum.}---It is known that the 6D  two-derivative supergravities admit a 
supersymmetric
${\rm AdS}_3\times {\rm S}^3$ solution where
the nonvanishing fields are given by
\bea
&ds^2_6=\varrho^2(ds^2_{{\rm AdS}_3}+ds^2_{{\rm S}^3}),\non\\
&H_{3}=2\varrho^{-1}(\Omega_{{\rm AdS}_3}-\Omega_{{\rm S}^3}),~~~~~~
L=1,
\label{ads3sl}
\eea
where $ds^2_{{\rm AdS}_3}$ and $ds^2_{{\rm S}^3}$ refer to the metrics on the unit radius 
of ${\rm AdS}_3$ and ${\rm S}^3$,
respectively, $\Omega_{{\rm AdS}_3}$ and $\Omega_{{\rm S}^3}$ denote their volume forms,
and $\varrho$ is the constant curvature radius. 
This solution arises as the near horizon limit of dyonic strings.
Similar to the ${\rm AdS}_5\times {\rm S}^5$ solution in type IIB \cite{Maldacena:1997re},
the ${\rm AdS}_3\times {\rm S}^3$ metric has vanishing Weyl tensor and scalar curvature \cite{Duff:1995yh}.
One can show that the ${\rm AdS}_3\times {\rm S}^3$ solution in eq.\,\eqref{ads3sl}
is also a solution of the Einstein-Gauss-Bonnet supergravity.
The spectrum of the two-derivative theory \eqref{eha} has been studied before in various works
\cite{deBoer:1998kjm,Deger:1998nm, Nicolai:2003ux}.
It contains only the short multiplets of ${\rm SU}(1,1|2)$ dressed by the irreducible representations of the extra
${\rm SL}(2,\mathbb{R})\times {\rm SU}(2)$, since the total
isometry group associated with the 
supersymmetric ${\rm AdS}_3\times {\rm S}^3$ vacuum is
${\rm SU}(1,1|2)\times {\rm SL}(2,\mathbb{R})\times {\rm SU}(2)$.
A short multiplet of ${\rm SU}(1,1|2)$ has the structure
\be
(h,j)\oplus2\times (h+\ft12,j-\ft12)\oplus(h+1,j-1),\quad h=j,
\label{sm1}
\ee
where $h$ and $j$ label the representations of the ${\rm SL}(2,\mathbb{R})\times {\rm SU}(2)$ bosonic subgroup inside
${\rm SU}(1,1|2)$. The irreducible representations of the extra ${\rm SL}(2,\mathbb{R})\times {\rm SU}(2)$ group 
are labeled by
$(\bar{h},\bar{j})$. All together,  we use ${\rm DS}^{(\bar{h},\bar{j})}(h,j)_{\rm S}$ to denote a short multiplet of
${\rm SU}(1,1|2)\times {\rm SL}(2,\mathbb{R})\times {\rm SU}(2)$. 
The spectrum of the two-derivative theory \eqref{eha} consists of
eight infinite towers of short multiplets, each of which is labeled by an integer $\ell\geq0$
\bea
&& {\rm DS}^{(\ft{\ell+3}2,\ft{\ell+3}2)}\Big(\ft{\ell+1}2,\ft{\ell+1}2\Big)_{\rm S},\,
2\times {\rm DS}^{(\ft{\ell}2+2,\ft{\ell}2)}\Big(\ft{\ell+2}2,\ft{\ell+2}2\Big)_{\rm S},
\non\\
&& {\rm DS}^{(\ft{\ell+1}2,\ft{\ell+1}2)}\Big(\ft{\ell+3}2,\ft{\ell+3}2\Big)_{\rm S},\,
{\rm DS}^{(\ft{\ell}2+2,\ft{\ell}2)}\Big(\ft{\ell+4}2,\ft{\ell+4}2\Big)_{\rm S},
\non\\
&& 2\times {\rm DS}^{(\ft{\ell}2+1,\ft{\ell}2+1)}\Big(\ft{\ell+2}2,\ft{\ell+2}2\Big)_{\rm S},\,
{\rm DS}^{(\ft{\ell}2+2,\ft{\ell}2)}\Big(\ft{\ell}2,\ft{\ell}2\Big)_{\rm S}.
~~~~~~
\label{sm2}
\eea
At each level $\ell$, the spectrum contains 16+16 degrees of freedom which is the same as the spectrum around 
supersymmetric ${\rm Minkowski}_6$.

Generic ${\rm SU}(1,1|2)$ long multiplets have the form \cite{deBoer:1998kjm}
\bea
&(h,j)\oplus2\times (h+\ft12,j-\ft12)\oplus(h+1,j-1)\oplus\non\\
&(h+\ft12,j+\ft12)\oplus2\times (h+1,j)\oplus (h+\ft32,j-\ft12).
~~~
\eea
Our analysis shows that the linearized spectrum around ${\rm AdS}_3\times {\rm S}^3$ of the action \eqref{GB-0}
contains, in addition to the eight
infinite towers of short multiplets \eqref{sm2}, four long multiplets whose AdS energies are independent of the
Kaluza-Klein (KK) level $\ell$, with $\ell\ge0$,
\bea
&&{\rm DS}^{(\frac{\varrho^2}{\a'},\frac{\ell}2)}\Big(\ft{\varrho^2}{\a'}+\ft12,\ft{\ell+1}2\Big)_{\rm L},~
 {\rm DS}^{(1-\ft{\varrho^2}{\a'},\ft{\ell}2)}\Big(-\ft{\varrho^2}{\a'}-{1},\ft{\ell}2\Big)_{\rm L},\non\\
&&{\rm DS}^{(1-\ft{\varrho^2}{\a'},\ft{\ell}2)}\Big(-\ft{\varrho^2}{\a'}-\ft12,\ft{\ell+1}2\Big)_{\rm L},
~{\rm DS}^{(\ft{\varrho^2}{\a'},\ft{\ell}2)}\Big(\ft{\varrho^2}{\a'},\ft{\ell}2\Big)_{\rm L},
\eea
where we have set $\kappa^2=1$ in \eqref{GB-0}.
It should be emphasized that the auxiliary vector $V^{ij}_{\m}$
has an important role in the analysis of the spectrum around
${\rm AdS}_3\times {\rm S}^3$.
Its KK massive states get reorganized into different multiplets. This would be impossible if one were 
considering the on-shell Gauss-Bonnet invariant \eqref{OnShellGB} where $V^{ij}_{\m}=0$.
The fact that all the states in our previous analysis fit into multiplets of 
${\rm SU}(1,1|2)\times{\rm SL}(2,\mathbb{R})\times {\rm SU}(2)$
provides a further consistency check of the supersymmetric invariance of the off-shell GB action \eqref{GBA}.

Unitarity requires $\bar{h}>0$ and $h>j\ge0$; thus, $\a'>0$ implies only the first and fourth series 
of the long multiplets can be
unitary, with the restriction $\ell\le\ft{\varrho^2}{\a'}$. However, unitarity of the representations 
carried by the linearized
modes means only that these modes are not tachyonic. Unitarity of the modes requires 
the Hamiltonian to be bounded from below.
In fact, the $B\wedge R\wedge R$ and $B\wedge F\wedge F$ terms in the GB action induce  Chern-Simons couplings 
in 3D via compactification on S$^3$,
and an Ostrogradsky-type analysis in the manner of \cite{Li:2008dq} shows that the Hamiltonian of the massive modes carrying unitary
representations are in fact unbounded from below. 
These nontachyonic ghostlike modes may encode the information of the
massive string states propagating in the supersymmetric AdS$_3\times$ S$^3\times$ K3(T$^4$) 
target space. The remaining modes
carrying nonunitary representations are tachyonic and nonrenormalizable. They may be removed from the spectrum by
imposing proper boundary conditions.

\textit{Outlook.}---The new GB invariant \eqref{GBA},  together with the Riemann \cite{BSS1,BSS2,Nishino:1986da,BR} and 
scalar curvature squared  \cite{OzkanThesis} combinations, allows one to construct all off-shell $\cN = (1, 0)$ 
curvature squared invariants in six dimensions.
These results give the opportunity 
to extend known supergravity-matter models by including  four-derivative  invariants. 
By having full control of the off-shell supersymmetry transformations,
one could determine whether BPS solutions of the two-derivative theory are solutions of the four-derivative ones.

In this letter, we 
studied the ${\rm AdS}_3\times {\rm S}^3$ solution arising from the near horizon limit of the dyonic string. 
However, we have not checked whether the full dyonic string, 
interpolating between the ${\rm AdS}_3\times {\rm S}^3$ and Minkowski$_6$, 
is unmodified by the GB invariant.
It will be interesting to explore the full solution in the Einstein-Gauss-Bonnet supergravity, 
from which one can also extract the central charge of the dual 2D SCFT. 

Compactifications to 4D of the GB invariant are also of interest. For instance,
the string-string-string duality observed in the 4D STU model 
can be extended to include the higher-derivative corrections by 
reducing the the 6D ${\cal N}=(1,0)$ Einstein-Gauss-Bonnet supergravity on a 2-torus \cite{Duff:1995sm}.

\textit{Acknowledgements.}---We are grateful to D.~Butter for discussions and collaboration on related projects. 
We also thank D.~Butter, F.~F.~Gautason and S.~Theisen for feedback and comments on the manuscript. 
JN acknowledges support from GIF,  the German-Israeli Foundation for
Scientific Research and Development. YP is supported by the Alexander von Humboldt fellowship. 
The work of GT-M was supported by the Interuniversity
Attraction Poles Programme initiated by the Belgian Science Policy (P7/37)
and in part by COST Action MP1210. 
The work of MO is supported in part by a Marie Curie Cofund Fellowship (No.116C028).


\begin{thebibliography}{2}

	
\bibitem{Green:1984bx}
M.~B.~Green, J.~H.~Schwarz and P.~C.~West,
Nucl.\ Phys.\ B {\bf 254}, 327 (1985).	


\bibitem{Walton:1987bu}
M.~A.~Walton,
Phys.\ Rev.\ D {\bf 37} (1988) 377.

		
\bibitem{Nishino:1986dc-1997ff}
H.~Nishino and E.~Sezgin,
Nucl.\ Phys.\ B {\bf 278}, 353 (1986);
Nucl.\ Phys.\ B {\bf 505}, 497 (1997).


\bibitem{Salam:1984cj}
A.~Salam and E.~Sezgin,
Phys.\ Lett.\  {\bf 147B}, 47 (1984).


\bibitem{Cvetic:2003xr}
M.~Cvetic, G.~W.~Gibbons and C.~N.~Pope,
Nucl.\ Phys.\ B {\bf 677} (2004) 164.


\bibitem{Gibbons:2003di}
G.~W.~Gibbons, R.~Gueven and C.~N.~Pope,
Phys.\ Lett.\ B {\bf 595}, 498 (2004).


\bibitem{Maeda:1984gq-1985es}
K.-i.~Maeda and H.~Nishino,
Phys.\ Lett.\  {\bf 154B}, 358 (1985);
Phys.\ Lett.\  {\bf 158B}, 381 (1985).


\bibitem{Gibbons:1986xp}
G.~W.~Gibbons and P.~K.~Townsend,
Nucl.\ Phys.\ B {\bf 282}, 610 (1987).


\bibitem{Halliwell:1986bs}
J.~J.~Halliwell,
Nucl.\ Phys.\ B {\bf 286}, 729 (1987).


\bibitem{Aghababaie:2002be}
Y.~Aghababaie, C.~P.~Burgess, S.~L.~Parameswaran and F.~Quevedo,
JHEP {\bf 0303}, 032 (2003).


\bibitem{deBoer:1998kjm}
J.~de Boer,
Nucl.\ Phys.\ B {\bf 548}, 139 (1999).


\bibitem{David:2002wn}
J.~R.~David, G.~Mandal and S.~R.~Wadia,
Phys.\ Rept.\  {\bf 369} (2002) 549.


\bibitem{Zwiebach:1985uq}
B.~Zwiebach,
Phys.\ Lett.\  {\bf 156B} (1985) 315.


\bibitem{Deser:1986xr}
S.~Deser and A.~N.~Redlich,
Phys.\ Lett.\ B {\bf 176} (1986) 350
Erratum: [Phys.\ Lett.\ B {\bf 186} (1987) 461].

	
\bibitem{Antoniadis:1997eg}
  I.~Antoniadis, S.~Ferrara, R.~Minasian and K.~S.~Narain,
  Nucl.\ Phys.\ B {\bf 507} (1997) 571.


\bibitem{Antoniadis:2003sw}
 I.~Antoniadis, R.~Minasian, S.~Theisen and P.~Vanhove,
  Class.\ Quant.\ Grav.\  {\bf 20} (2003) 5079.


\bibitem{Liu:2013dna}
J.~T.~Liu and R.~Minasian,
Nucl.\ Phys.\ B {\bf 874} (2013) 413.


\bibitem{N=1-GB}
S.~Cecotti, S.~Ferrara, L.~Girardello and M.~Porrati, Phys.\ Lett.\  {\bf 164B} (1985) 46;
S.~Theisen, Nucl.\ Phys.\ B {\bf 263}, 687 (1986);
I.~L.~Buchbinder and S.~M.~Kuzenko, 
Preprint No. 35 of the Tomsk Branch of the Siberian Division of the USSR Academy of Science (November, 1985),
Nucl.\ Phys.\ B {\bf 308}, 162 (1988);
S.~Cecotti, S.~Ferrara, L.~Girardello, M.~Porrati and A.~Pasquinucci,
Phys.\ Rev.\ D {\bf 33} (1986) 2504;
S.~Ferrara, S.~Sabharwal and M.~Villasante,
Phys.\ Lett.\ B {\bf 205}, 302 (1988);
S.~Ferrara and M.~Villasante, J.\ Math.\ Phys.\  {\bf 30}, 104 (1989);
R.~Le Du, Eur.\ Phys.\ J.\ C {\bf 5}, 181 (1998).


\bibitem{Butter:2013lta}
D.~Butter, B.~de Wit, S.~M.~Kuzenko and I.~Lodato,
JHEP {\bf 1312} (2013) 062.


\bibitem{OP131}
M.~Ozkan and Y.~Pang,
JHEP {\bf 1303} (2013) 158
Erratum: [JHEP {\bf 1307} (2013) 152].


\bibitem{OP132}
M.~Ozkan and Y.~Pang,
JHEP {\bf 1308}, 042 (2013).


\bibitem{Butter:2014xxa}
D.~Butter, S.~M.~Kuzenko, J.~Novak and G.~Tartaglino-Mazzucchelli,
JHEP {\bf 1502} (2015) 111.


\bibitem{BSS1}
E.~Bergshoeff, A.~Salam and E.~Sezgin,
Phys.\ Lett.\ B {\bf 173}, 73 (1986).


\bibitem{BSS2}
E.~Bergshoeff, A.~Salam and E.~Sezgin,
Nucl.\ Phys.\ B {\bf 279}, 659 (1987).


\bibitem{Nishino:1986da}
H.~Nishino and S.~J.~Gates, Jr.,
Phys.\ Lett.\ B {\bf 173} (1986) 417.


\bibitem{BR}
E.~Bergshoeff and M.~Rakowski,
Phys.\ Lett.\ B {\bf 191}, 399 (1987).


\bibitem{BSVanP}
E.~Bergshoeff, E.~Sezgin and A.~Van Proeyen,
Nucl.\ Phys.\ B {\bf 264}, 653 (1986),
Erratum: [Nucl.\ Phys.\ B {\bf 598}, 667 (2001)].


\bibitem{BKNT16}
D.~Butter, S.~M.~Kuzenko, J.~Novak and S.~Theisen,
JHEP {\bf 1612}, 072 (2016).


\bibitem{BNT-M17}
D.~Butter, J.~Novak and G.~Tartaglino-Mazzucchelli,
  JHEP {\bf 1705}, 133 (2017).

\bibitem{Romans:1985tw}
L.~J.~Romans,
Nucl.\ Phys.\ B {\bf 269} (1986) 691.


\bibitem{Coomans:2011ih}
F.~Coomans and A.~Van Proeyen,
JHEP {\bf 1102}, 049 (2011).


\bibitem{Bergshoeff:2012ax}
E.~Bergshoeff, F.~Coomans, E.~Sezgin and A.~Van Proeyen,
JHEP {\bf 1207} (2012) 011.


\bibitem{BF4}
P.~Breitenlohner and M.~F.~Sohnius,
Nucl.\ Phys.\ B {\bf 165}, 483 (1980);
B.~de Wit, J.~W.~van Holten and A.~Van Proeyen,
Phys.\ Lett.\  {\bf 95B}, 51 (1980);
B.~de Wit, R.~Philippe and A.~Van Proeyen,
Nucl.\ Phys.\ B {\bf 219}, 143 (1983).


\bibitem{BF5}
M.~Zucker,
JHEP {\bf 0008}, 016 (2000);
T.~Kugo and K.~Ohashi,
Prog.\ Theor.\ Phys.\  {\bf 104}, 835 (2000).


\bibitem{BF3}
S.~M.~Kuzenko and J.~Novak,
JHEP {\bf 1405}, 093 (2014).


\bibitem{BNOPT-M17}
D.~Butter, J.~Novak, M.~Ozkan, Y.~Pang  and G.~Tartaglino-Mazzucchelli,
\emph{work in progress}.


\bibitem{AriasLinchRidgway}
C.~Arias, W.~D.~Linch, III and A.~K.~Ridgway,
JHEP {\bf 1605}, 016 (2016).


\bibitem{HoweSierraTownsend}
P.~S.~Howe, G.~Sierra and P.~K.~Townsend,
Nucl.\ Phys.\ B {\bf 221}, 331 (1983).


\bibitem{Maldacena:1997re}
J.~M.~Maldacena,
Int.\ J.\ Theor.\ Phys.\  {\bf 38}, 1113 (1999)
[Adv.\ Theor.\ Math.\ Phys.\  {\bf 2}, 231 (1998)].

		
\bibitem{Duff:1995yh}
M.~J.~Duff, S.~Ferrara, R.~R.~Khuri and J.~Rahmfeld,
Phys.\ Lett.\ B {\bf 356}, 479 (1995).


\bibitem{Deger:1998nm}
S.~Deger, A.~Kaya, E.~Sezgin and P.~Sundell,
Nucl.\ Phys.\ B {\bf 536} (1998) 110.


\bibitem{Nicolai:2003ux}
H.~Nicolai and H.~Samtleben,
JHEP {\bf 0309} (2003) 036.


\bibitem{Li:2008dq}
W.~Li, W.~Song and A.~Strominger,
JHEP {\bf 0804} (2008) 082.


\bibitem{OzkanThesis}
M.~Ozkan,
``Supersymmetric curvature squared invariants in five and six dimensions,''
\href{http://oaktrust.library.tamu.edu/bitstream/handle/1969.1/151223/OZKAN-DISSERTATION-2013.pdf?sequence=1}
{PhD Thesis}, Texas A\&M University, 2013.


\bibitem{Duff:1995sm}
M.~J.~Duff, J.~T.~Liu and J.~Rahmfeld,
Nucl.\ Phys.\ B {\bf 459} (1996) 125.



\end{thebibliography}
\end{document}